\documentclass[twocolumn,english,notitlepage,superscriptaddress,nofootinbib]{revtex4-1}
\usepackage[T1]{fontenc}
\usepackage[latin9]{inputenc}
\usepackage{float}
\usepackage{amsmath}
\usepackage{amssymb}
\usepackage{graphicx}
\usepackage{wasysym}
\usepackage{lipsum}
\usepackage{color}
\makeatletter

\begin{document}

\title{Hyperpolarisation Dynamics: Asymptotic Polarisation}

\author{O.T. Whaites} \affiliation{Department of Physics and Astronomy, University College London,
Gower Street, London WC1E 6BT, United Kingdom}

\author{T.S. Monteiro} \affiliation{Department of Physics and Astronomy, University College London,
Gower Street, London WC1E 6BT, United Kingdom}

\begin{abstract} 
For applications of solid state quantum computing and quantum simulations, high fidelity initialisation of thermally mixed electronic and nuclear spin qubits is essential. Whereas electronic spins can readily be initialised optically to high fidelity, initialisation of the nuclear spins requires alternative approaches, such as dynamic nuclear polarisation (DNP) via the electronic spin. Pulse-based DNP methods, such as PulsePol, are already widely utilised. By means of repeated application of PulsePol sequences, interspersed with re-initialisation of the  electronic spin, high levels of nuclear polarisation - termed hyperpolarisation-  have been achieved. From theoretical analysis of these protocols perfect nuclear initialisation is expected; however, in practice, saturation below $\sim$ 95$\%$ is seen in experiment. We develop an analytical model to describe hyperpolarisation dynamics, predicting non-maximal nuclear polarisation saturation in the asymptotic limit for realistic nuclear spin clusters. We argue that perfect initialisation of a typical nuclear cluster using this method may not, in general, be possible even with an arbitrarily large number of repetitions.

\end{abstract}

\maketitle

\section{Introduction}

Solid state spin defects such as nitrogen vacancy (NV) centers in diamond \cite{Degen2014} show great promise in the development of a raft of quantum technologies \cite{Jelezko2004,Maze2008a, Bala2008,sivac1,sivac2}. Said defects have been extensively used for the control of nuclear spins, demonstrating important advancements towards quantum computing and nanoscale NMR \cite{Aslam2017}. Consequently, significant interest has been directed towards developing robust and efficient protocols for nuclear control via electronic central spins.  

Originally designed to decouple a central spin from the environment, 
pulsed based dynamical decoupling (DD) methods such as CPMG have demonstrated improved spin state coherence times \cite{DD, BarGill}. However, dramatic dips in coherence were soon observed, attributed to entanglement between the central spin and proximate nuclear spins at resonant DD pulse spacing. This has opened the way to nanoscale sensing of single nuclear spins \cite{Kolkowitz2012, Tam2012,Abobeih2019,Stolpe2023} as well as  the control of surrounding nuclear spins, extending applications to nuclear long-lived memory \cite{Tam2014, Tam2019}.   

\begin{figure}[ht!] 
	\includegraphics[width = 3.6in]{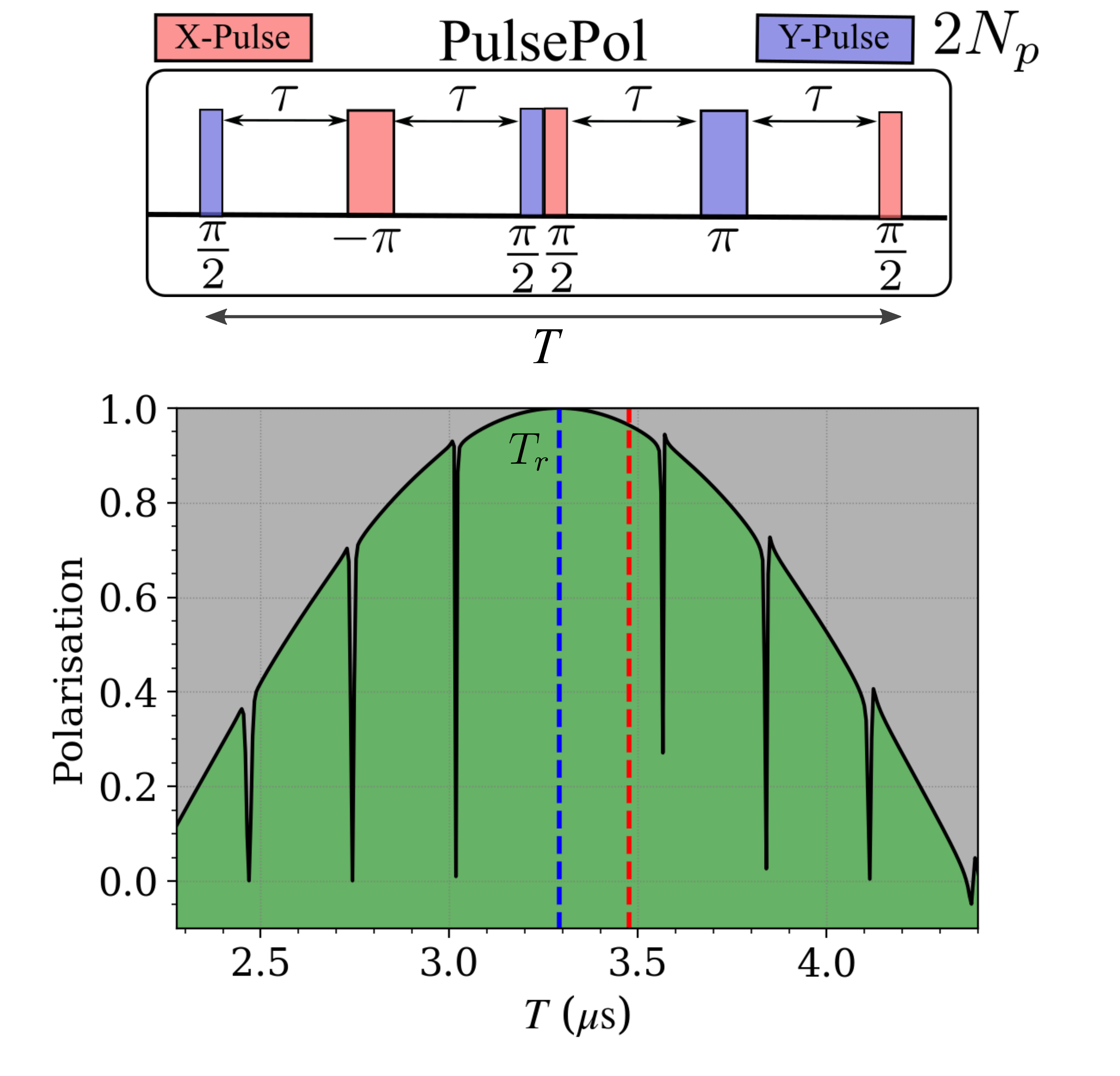}
	\caption{Illustration of below maximum saturation of nuclear $^{13}\mathrm{C}$ spin polarisation via off-resonant PulsePol using a NV center. The \textbf{upper panel} defines the pulse protocol studied here, PulsePol, which is incident on an electronic spin register. In the \textbf{lower panel} we investigate the asymptotic nuclear polarisation envelope with $T$. After many repetitions ($R =$ 200,000 repetitions) simulations of nuclear polarisation reach an asymptotic limit for all $T$ and maximal polarisation is achieved at resonant periodicity $T_r$, blue dashed line. However, this choice may not be possible or the most efficient. In fact, for a nuclear spin with modest detuning $A_z/2\pi \simeq 30 \,\mathrm{kHz}$, the na\"ive choice of matching the periodicity with the nuclear Larmor frequency, the red dashed line, only achieves $\simeq 95\%$ polarisation in the asymptotic limit.}
	\label{Fig1}
\end{figure}

Applications in quantum computing, such as qubit state storage in nuclear-spin memories, require initialisation into high fidelity pure states. In contrast nuclear spins are often in thermally mixed states that are typically very weakly polarised, with spin-$1/2$ impurities close to 50:50 ratios of up and down spin states. Thus efficient and robust protocols are needed that can initialise ensembles of nuclear spins in states of very high statistical polarisation. 

Recent pulse protocols such as PulsePol and PolCPMG \cite{PulsePol, PulsePol1, PolCPMG} were designed to enable initialisation of nuclear spin states via dynamic nuclear polarisation (DNP) \cite{Goldman1978, Bajaj2003}. Such protocols have been shown to transfer the polarisation of an electronic spin to individual neighbouring nuclear spins. By means of repeated application of the protocols, interspersed with re-initialisation of the \textit{electronic} spin, hyperpolarisation of both proximate as well as external nuclear spin \textit{clusters} has been demonstrated with NV centers \cite{hypol,hypol1}.

Initialisation of multiple nuclear spins simultaneously, using hyperpolarisation methods, may be employed to prepare polarised clusters. A recent experiment demonstrated collective 'time-crystal'  dynamics of a  nuclear spin cluster that had been initialised with PulsePol with saturation at about 75\% statistical polarisation \cite{Randall2021}. However, for applications such as quantum information or quantum simulation much higher polarisations >99\% are generally required.  

Previous studies have addressed the role of dark states and polarisation blockade \cite{blockade} in hyperpolarisation, but, to our knowledge, the long-time hyperpolarisation dynamics of a cluster, including the asymptotic limits of statistical polarisation has not yet been analysed. Analytical modelling of this process may be key to understand the physical processes that limit experimentally achievable polarisation of spin ensembles and to propose strategies for overcoming this limit.

In this study we show that there is in fact an asymptotic nuclear polarisation envelope for hyperpolarisation; we derive a closed form expression for the envelope using a Markov chain model, highlighting that for some DNP parameters the transfer of polarisation not only slows, but reaches below maximal saturation in the asymptotic limit. We conjecture that this may account for the experimental observation that perfect polarisation of a spin cluster is unattainable in general, even if the DNP is repeated an arbitrarily large number of times.

\begin{figure}[ht!] 
	\includegraphics[width = 3.6in]{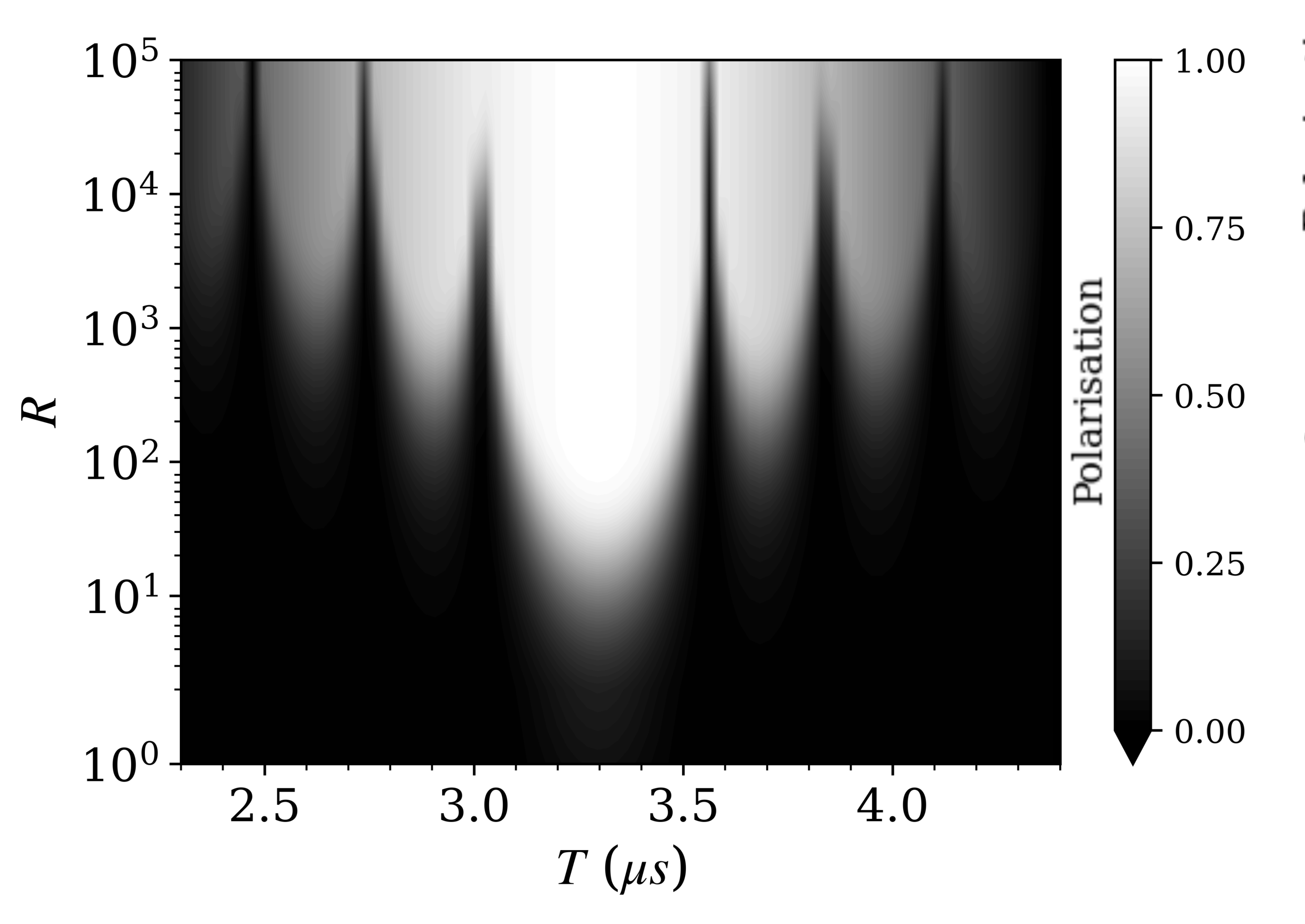}
	\caption{Simulation of polarisation convergence with $R$ to a saturated level. Colour map is the polarisation of nuclear spin, $\mathcal{P}_R$ with hyperfine couplings $(A_z,A_x)/2\pi \simeq (-50,9)\,\mathrm{ kHz}$ for PulsePol with $N_p = 4$ cycles. Convergence to the asymptotic envelope is reached after $R\sim 10^4$ repetitions. Saturated polarisation levels are increasingly below maximum level of $\mathcal{P}_R = 1$ the further $T$ is from resonant $T_r$.}
	\label{Fig2}
\end{figure}

\section{Methods}

We consider a electronic spin under pulsed microwave driving coupled to a single nuclear spin, such that the \emph{pure-dephasing} Hamiltonian is $\hat{H}(t) = \hat{H}_p(t) + \hat{H}_0$ where
\begin{equation}
\hat{H}_0 = \omega_L\hat{I}_z + \hat{S}_z(A_z\hat{I}_z + A_x\hat{I}_x)
\end{equation}
The operators $\hat{S}$ and $\hat{I}$ operators are in the electronic spin and nuclear spin state space respectively. $\omega_L$ is the nuclear Larmor frequency and $A_z,\,A_x$ are the hyperfine couplings relative to an electronic spin $z$-axis. The pulse Hamiltonian is dependent on the protocol, where for each $x$/$y$ pulse $\hat{H}_p(t) = \Omega_{x/y}(t)\hat{S}_{x/y}$ with microwave control field $\Omega_{x/y}$.

We study the DNP pulse protocol PulsePol, schematic in Fig.\ref{Fig1}. It is understood that when the pulse spacing is resonant with the nuclear precession frequency, the state of the electronic spin is transferred to a particular state of the nuclear spin. This transfer is dependent on the initial state of the electronic spin. The resonant pulse spacing for PulsePol is $\tau = \tau_r = k\pi/(4\omega_I)$ for harmonic $k$ (odd integer) where $\omega_I = \sqrt{(\omega_L - A_z/2)^2 + (A_x/2)^2}$ is the average nuclear precession frequency. At the resonant pulse spacing for the $k = 3$ harmonic, the effective Hamiltonian to first order in the Magnus expansion is a `flip-flop' Hamiltonian between the electronic spin and nuclear spin, $\hat{H} = g_3(\hat{S}_+\hat{I}_- + \hat{S}_-\hat{I}_+)$ with effective coupling  $g_3 = A_x(\sqrt{2} + 2)/6\pi$, a result key from \cite{PulsePol}.

Off-resonance PulsePol has been considered recently in \cite{blockade}. Where usually the Hamiltonian is transformed into the rotating frame of the nuclear spin, alternatively a reference frame close to this can be chosen, such that there is a nuclear-detuning $\delta = \omega_I - 3\pi/T$, for the $\textrm{3}^\mathrm{rd}$ resonance. As the frame chosen is close to the nuclear frame, this detuning is assumed to be small ($\delta \ll 1/T$) and usual perturbations are used. The average Hamiltonian in this frame is given to be
\begin{equation}
\hat{H}(T) = \delta(T)\hat{I}_z + g_3(\hat{S}_+\hat{I}_- + \hat{S}_-\hat{I}_+)
\label{1st_order_ham}
\end{equation}
the usual resonant PulsePol Hamiltonian with an extra nuclear term. In \cite{blockade} this was studied in the context of spin-pair dark states, although here we study this for the off-resonance polarisation of single spins.

Higher levels of polarisation can be reached by repeatedly applying polarisation protocols such as PulsePol interspersed with electronic spin re-initialisation. For this, coherent polarisation transfer between the electronic and nuclear spin is performed for $N_p$ cycles. Subsequently, the electronic spin is re-initialised into the ground state, ideally leaving the nuclear spin dynamics uninterrupted. This process is then repeated $R$ times until maximal levels of polarisation are reached.

\begin{figure*}[ht!] 
	\includegraphics[width = 6.5in]{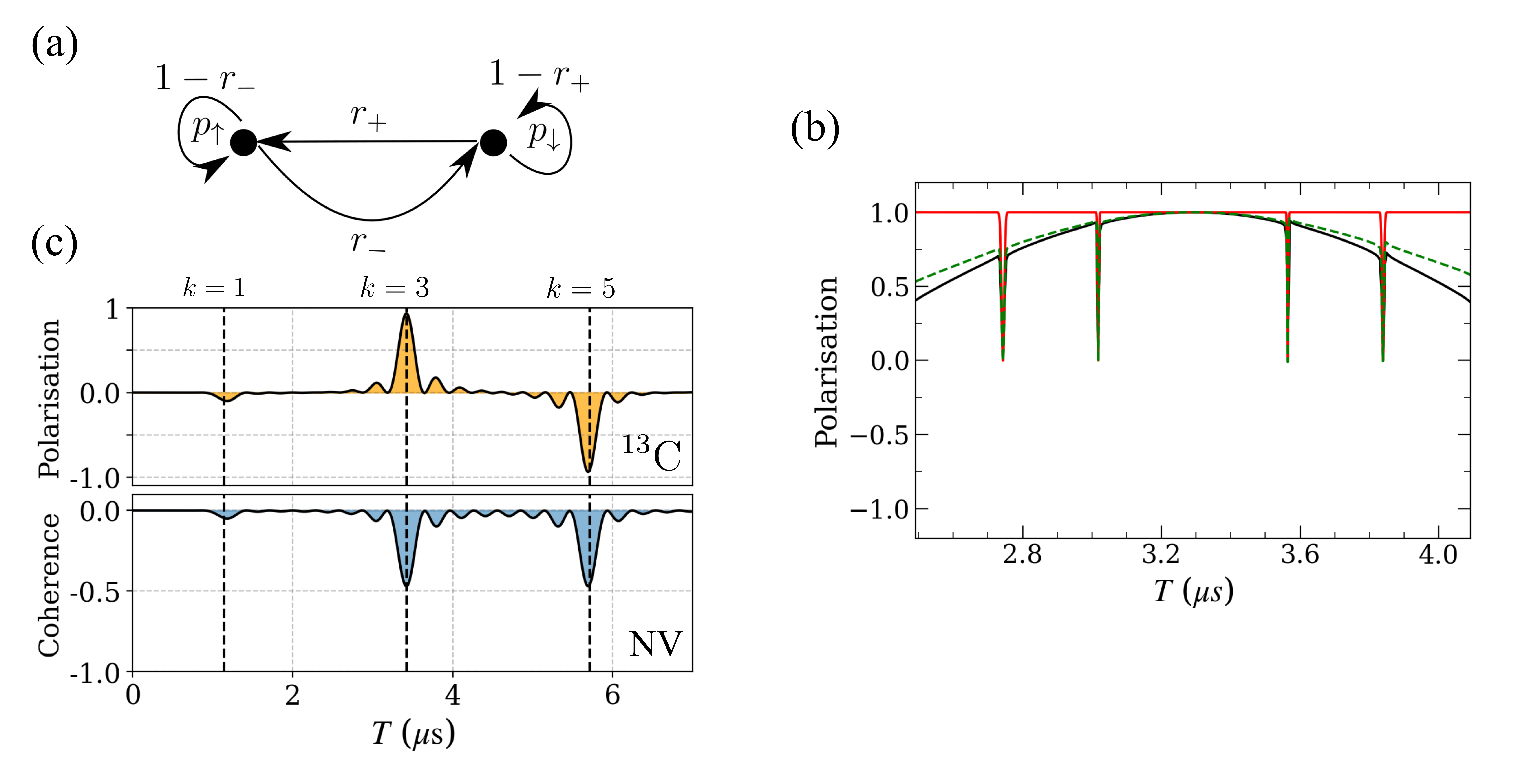}
	\caption{Comparisons of analytical Markov chain model and simulations. (a) shows a graphical representation of the Markov chain for repeated nuclear spin polarisation. Nuclear states are black nodes and directional transition probabilities are directional vertices. The left node and right node labelled $p_\uparrow,\, p_\downarrow$ represent the $|\uparrow\rangle$ and $|\downarrow\rangle$ states respectively. (b) Simulation (black dashed line) of the repeated polarisation of nuclear spin with hyperfine couplings $(A_z,A_x)/2\pi \simeq (-50,9)\,\mathrm{ kHz}$ for $R =$ 200,000 repetitions of the PulsePol protocol with $N_p = 4$ cycles. This is compared to Eq.\eqref{pol_sat}, an analytical Markov chain model. Transition probabilities derived using the first order Hamiltonian from Eq.\eqref{1st_order_ham} are shown as a red line and second order Hamiltonian Eq.\eqref{2nd_order_ham} are shown as a green line, with the latter capturing full simulations better. (c) Highlights the opposite polarisation direction for neighbouring harmonics $k = 1,5$ to $k = 3$ of PulsePol without repetitions. Simulations of harmonics are for a nuclear spin with hyperfine couplings $(A_z,A_x)/2\pi \simeq (-10,60)\,\mathrm{ kHz}$ and $N_p = 4$.}
	\label{Fig3}
\end{figure*}

Typically, inhomogeneous perpendicular couplings $A_x$ for each nuclear spin requires a unique number protocol cycles $N_p$ to achieve high levels of polarisation. However, introducing repetitions can yield universally acceptable polarisation for nuclear spins of all coupling strengths, given enough repetitions are made. Figure \ref{Fig2} illustrates the convergence of polarisation for a single nuclear spin with perpendicular coupling $A_x/2\pi \simeq 9\,\mathrm{kHz}$ with $R$ for a range of off-resonant $T$. Here is shown that polarisation saturation is reached for $R \sim 10^4$, at least for spins with perpendicular coupling $A_x/2\pi > 9\,\mathrm{kHz}$.

To analyse the repeated polarisation of a nuclear spin, we employ a Markov chain model. A Markovian process is a stochastic model which assumes no hysteresis. Explicitly, the future state of the system only depends on the current state, not any past trajectory. A Markov chain is a discrete-time Markovian process, in which the system has a probability of transitioning to another state at set time steps. The probability of being in a state $n$ at a $R^\mathrm{th}$ discrete time step, $t_R$, for a Markov chain is 
\begin{equation}
p(n,t_R) = \sum^{D}_m p(n,t_R | m,t_{R - 1})p(m,t_{R - 1})
\end{equation} 
where $D$ is the number of discrete states.

The vector of state probabilities, or the state distribution, at time $t_R$ defined as $\mathbf{p}_{t_R} = (...,p(n,t_R),p(n + 1,t_R),...)^T$ evolves according to $\mathbf{p}_{t_R} = Q\mathbf{p}_{t_{R-1}} = Q^R\mathbf{p}_{0}$ where the $Q$ is the $D\times D$ transition matrix with matrix elements $Q_{nm} = p(n,t_{R}|m,t_{R - 1})$ and $\mathbf{p}_{0}$ is the initial distribution of the system. If the process can be represented by a Markov chain with no terminal states, then the probability distribution can be decomposed into the basis of eigenvectors, $\mathbf{v}_n$ (with corresponding eigenvalues $\lambda_n$) of $Q$, where the distribution evolves as
\begin{equation}
\mathbf{p}(t_R) = \sum_n \lambda^R_n c_n(0)\mathbf{v}_n
\end{equation}
defining $c_n(0)$ as the expansion coefficients of the initial distribution in terms of the $Q$ eigenbasis. Additionally, one of the eigenvalues, $\lambda_n$ of $Q$ will be equal to unity, with all others within a unit disk in the complex plane $|\lambda_{m \neq n}| < 1$. Then, after a large amount of discrete time steps, all eigenvectors with non-unity eigenvalues will decay and the system will converge to a stationary distribution equal to the eigenstate $\mathbf{v}_n$.

\section{Results}

Full details of this work are given in the supplementary information. Here we summarise key results. The analytical framework presented here is fairly general, however we apply it in the context to a NV center in diamond coupled to a single $^{13}\mathrm{C}$ spin-$1/2$ impurity. The spin-1 NV center is taken to be in the $\{|0\rangle,|-1\rangle\}$ pseudospin-$1/2$ subspace.

We model the repeated polarisation of a nuclear spin via an electronic spin as a Markov chain. As the electronic spin is continuously re-initialised, any quantum correlations with the nuclear spin are destroyed and hence this process is assumed to be a semi-classical process an a Markov chain is appropriate (see supplementary information). Therefore the polarisation of a single nuclear spin is a two state ($|\uparrow\rangle,|\downarrow\rangle$) Markov chain with a probability distribution vector $\mathbf{p}(t_R) = (p_\uparrow(t_R),p_\downarrow(t_R))^T$.

A nuclear spin initially in a 50:50 thermally mixed state has a probability distribution $\mathbf{p}(0) = (1/2,1/2)^T$. Within each repetition the nuclear spin has a probability of transitioning into another state. We denote the transition probability from $|\downarrow\rangle \to |\uparrow\rangle$ by $r_+$ and from $|\uparrow\rangle \to |\downarrow\rangle$ by $r_-$. These probabilities will be calculated using the Hamiltonian of the system. A graphical representation of this Markov chain is shown in Fig.\ref{Fig3}(a).

Polarisation of this distribution is defined as $\mathcal{P}_R = \sum_{n}(\sigma_z \mathbf{p}(t_R))$ where $\sigma_z$ is the $z$-directional Pauli matrix. For the Markov chain in Fig.\ref{Fig3}, we find the stationary distribution has a polarisation of

\begin{equation}
\mathcal{P}_\infty(N_p T) = \lim_{R \rightarrow \infty} \mathcal{P}_R(N_p T) = \frac{r_+ - r_-}{r_+ + r_-}
\label{pol_sat}
\end{equation}
where it is implied that $r_\pm \equiv r_\pm(N_p T)$.

The off-resonance polarisation of a single nuclear spin using PulsePol is described by the Hamiltonian in Eq.\eqref{1st_order_ham}. After $N_p$ cycles, we find the transition probabilities to be 
\begin{equation}
r_+ = \left(\frac{g_3}{\Omega_r}\right)^2\sin^2\left(\frac{\Omega_r N_p T}{2}\right)
\end{equation}
and $r_- = 0$, where the generalised Rabi frequency $\Omega_r = \sqrt{\delta^2 + 4g_3^2}$. Figure \ref{Fig3}(b) shows the polarisation envelope of a Markov chain with these transition probabilities. As the transition probability $r_- = 0$, the $|\uparrow\rangle$ state is a terminal state, the asymptotic envelope in Eq.\eqref{pol_sat} is, in general, $\mathcal{P}_\infty = 1$. Exceptions to this are present, as there are sharp dips where $\mathcal{P}_\infty = 0$. These are known as side-dips studied in the supplementary information, for which the $r_+ = 0$ as well as $r_- = 0$ and the dynamics is trivial. 

The time averaged PulsePol Hamiltonian, even with the addition of nuclear detuning, fails to capture the asymptotic envelope in simulations, where saturation below maximal polarisation is seen off-resonance.  

Harmonics in PulsePol alternate in their polarisation direction. That is, whilst the harmonic of interest $k = 3$ polarises the nuclear spin into the $|\uparrow\rangle$ state, for the same initial electronic spin state, the neighbouring $k = 1,5$ harmonics polarise into the opposite $|\downarrow\rangle$ state. This is demonstrated in Fig.\ref{Fig3}(c). Any overlap of these harmonics may cause non-optimal polarisation, an effect which is commonly neglected. However, to first order the contribution to the Hamiltonian from neighbouring harmonics is exactly zero, yielding in the same Hamiltonian in Eq.\eqref{1st_order_ham}. We consider higher order terms.

It is assumed that the perpendicular coupling, $A_x \ll 1/T$, and so higher order terms in the Magnus expansion are negligible. However, the same assumption cannot be made for terms involving $\delta$, as this increases for $T$ further away from the resonance $T_r$. By retaining all terms $\propto A_x \delta$, we find the second order Hamiltonian to be

\begin{equation}
\begin{split}
\hat{H}(T) \simeq \delta(T)\hat{I}_z &+ g_+(T)(\hat{S}_+\hat{I}_- + \hat{S}_-\hat{I}_+) \\ &+ g_-(T)(\hat{S}_+\hat{I}_+ + \hat{S}_-\hat{I}_-)
\end{split}
\label{2nd_order_ham}
\end{equation}
where for the $3^\mathrm{rd}$ harmonic $g_+(T) \simeq g_3 + \delta T(2g_1 - g_5)/8\pi$ and $g_-(T) \simeq \delta T(g_3 + 3g_1 + 3g_2 )/6\pi$ with effective couplings for the $1^\mathrm{st}$ and $5^\mathrm{th}$ harmonic $g_1 =  A_x(2 - \sqrt{2})/2\pi$ and $g_5 =  A_x(\sqrt{2} + 2)/10\pi$ respectively. Crucially, terms proportional to $g_-$ allow for `de-polarisation', or non-zero $r_-$.

We find the transition probabilities for this second order Hamiltonian to be 
\begin{equation}
r_{\pm} = \left(\frac{g_\pm}{\Omega_\pm}\right)^2\sin^2\left(\frac{\Omega_\pm N_p T}{2}\right)
\label{transition_2nd}
\end{equation}
where the generalised Rabi- frequencies are $\Omega_\pm = \sqrt{\delta^2 + 4g_\pm^2}$. Figure \ref{Fig3} shows the asymptotic envelope in Eq.\eqref{pol_sat} for the second order transition probabilities. The envelope exhibits non-maximal saturation for off-resonant $T$, as is present in the simulation. Agreement with simulation is excellent for smaller detuning $\delta$, but is increasingly worse as this grows. Better agreement may be achieved by considering higher order terms of order $A_x \delta^m$ for $m > 1$.

\begin{figure}[t!] 
	\includegraphics[width = 3.6in]{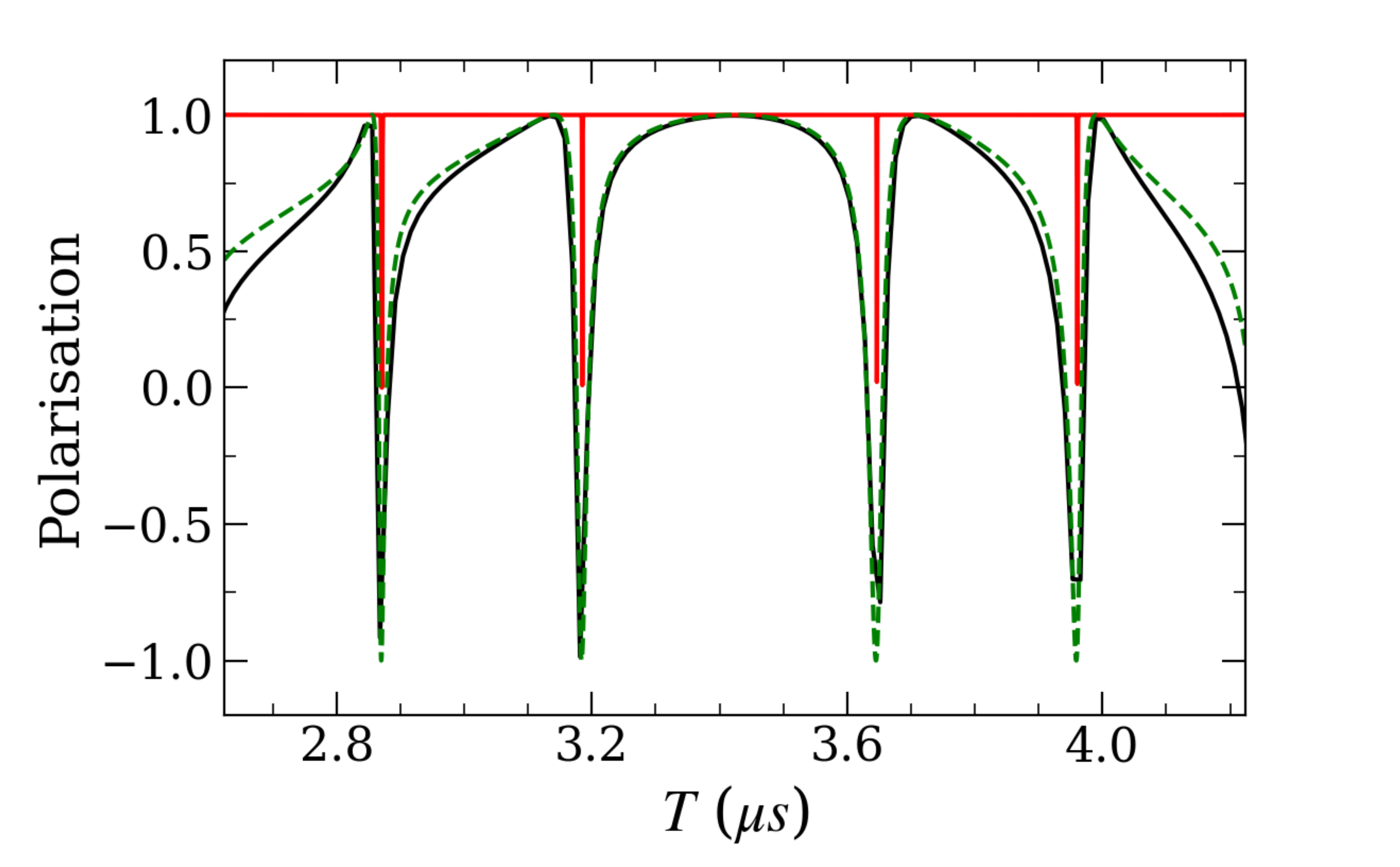}
	\caption{Illustrates the asymptotic polarisation envelope for a nuclear spin with a stronger perpendicular coupling. Simulations and Markov models are for a nuclear spin with hyperfine coupling $(A_z,A_x)/2\pi \simeq (-10,60)\,\mathrm{ kHz}$ for $R =$ 200,000 repetitions of $N_p = 4$ cycles of PulsePol. For both simulation (black dashed) and Markov chain models with second order transition probabilities (green) have much broader side dips, where opposite polarisation $\mathcal{P}_\infty = -1$ is reached rather than $\mathcal{P}_\infty = 0$. Additionally, maximal polarisation is achieved away from the resonance, proximate to each side dip.}
	\label{Fig4}
\end{figure}

Fig.\ref{Fig4} shows the polarisation of a more strongly coupled spin after $R =$ 200,000 repetitions. This demonstrates that the asymptotic envelope is also dependent on the strength of the perpendicular hyperfine coupling $A_x$. Side dips are seen to be much broader and reach opposite polarisation of $\mathcal{P}_\infty = -1$. Additionally, off-resonance maximal polarisation $\mathcal{P}_\infty = 1$ is achieved in this proximity.

This is due to condition for $r_- = 0$ and $r_+- = 0$ no longer being degenerate, as this depends on the strength of $A_x$. Therefore when $r_+ = 0$, $r_-$ allows for maximal opposite polarisation and vice-versa.

\section{Discussion}

Nuclear spins with a detuning of $A_z/2\pi \sim 30\,\mathrm{kHz}$ will have a loss of $\sim 5\%$ fidelity during initialisation for DNP at the standard Larmor frequency for a magnetic field of $B_0\simeq 400 \,\mathrm{G}$ and perpendicular coupling $A_x/2\pi \simeq 10\,\mathrm{kHz}$. For stronger coupled spins this loss in fidelity may be more drastic due broader side dips, as seen in Fig.\ref{Fig4}. Ideally, the highest fidelities when simultaneously initialising multiple nuclear spins are achieved for clusters with similar resonances and hence similar $A_z$. However in this regime, one may risk multi-spin polarisation blocking effects such as the formation of dark states and polarisation blockades \cite{blockade}.

In fact the statistical polarisation limit in Eq.\eqref{pol_sat} is more general. If the periodicity is chosen such that the asymptotic polarisation for a particular nuclear spin is driven to $\mathcal{P}_\infty(T_1) = 1$, by then changing the periodicity, the nuclear spin does not remain in that statistical mixture. Instead it will converge to a new statistical polarisation likely with $P_\infty(T_2) \neq 1$. Hence, two nuclear spins with vastly different resonances $T_1$ and $T_2$ may not be simultaneously fully initialised using this DNP scheme.

Equally, even for DNP protocols at optimal $T_r$ with nuclear detuning $\delta(T_r) = 0$, where maximal polarisation is expected in the asymptotic limit, experimental fluctuations of the applied magnetic field could result in polarisation saturation below maximum. Changing magnetic fields would alter the expected value of $T_r$ throughout the polarisation process, and so remaining at the original resonant $T$ would introduce non-zero nuclear detuning and depolarisation. In this case, the asymptotic polarisation limit would largely depend on the magnitude and timescale of the magnetic field fluctuations.

Cluster polarisation may be potentially improved by utilising adiabatic methods of state transfer, such as those in \cite{AdPulse, Villazon2021}. Here, there is no issue with choice of pulse spacing, where the limit of polarisation is set by how adiabatic the sweep is and hence the coherence times of the NV. This could provide a smoother transition to maximally polarised nuclear states.

\section{Conclusions}

We investigate the asymptotic envelope of nuclear spin polarisation via the DNP using a NV center, explicitly the pulse protocol PulsePol. We argue that the understanding of off-resonance polarisation is essential and show that truncating at low order terms does not fully replicate the asymptotic envelope.

{\bf Acknowledgements} 
We would like to thank Tim Taminiau, Joe Randall and others at Delft University of Technology who drew our attention to the problem which this study seeks to address, for useful discussions and insights regarding their experiment. Oliver Whaites acknowledges support from an EPSRC DTP grant.

\newpage
\onecolumngrid

\end{document}